\documentclass[preprint,preprintnumbers,amsmath,amssymb]{revtex4}

\usepackage{graphicx}% Include figure files
\usepackage{dcolumn}% Align table columns on decimal point
\usepackage{bm}% bold math

\begin{document}

\title{ENERGY AND MOMENTUM OF A STATIONARY BEAM OF LIGHT\\}

\author{Thomas Bringley}
 \email{ttb2@duke.edu}
\affiliation{Physics and Mathematics Departments, Duke University
\\Physics Bldg., Science Dr., Box 90305 \\Durham, NC 27708,USA}
\date{\today}

\begin{abstract}
The energy-momentum complexes of Einstein, Landau-Lifshitz,
Papapetrou, and Weinberg give the same and meaningful results for
the energy and momentum of the Bonnor spacetime describing the
gravitational field of a stationary beam of light.  The results
support the Cooperstock hypothesis.
\end{abstract}

\maketitle

\section{\label{sec:level1}Introduction:\protect\\}

    Since Einstein proposed the General Theory of Relativity,
relativists have not been able to agree upon a definition of the
energy-momentum distribution associated with the gravitational
field (see Ref. \cite{debate} and references therein). In 1990, H.
Bondi\cite{Bondi} argued that General Relativity does not permit a
non-localizable form of energy, so, in principle, we should expect
to be able to find an acceptable definition.

    Energy and momentum density are usually defined by a second
rank tensor $T_{i}^{k}$.  The conservation of energy and momentum
are described by the requirement that the tensor's divergence is
zero. However, in General Relativity, the partial derivative in
the usual conservation equation $T_{i,k}^{\ k}=0$ is replaced by a
covariant derivative. $T_{i}^{\ k}$ then represents the energy and
momentum of matter and all non-gravitational fields and no longer
satisfies $T_{i\ ,k}^{\ k}=0$. A contribution from the
gravitational field must be added to obtain an energy-momentum
expression with zero divergence. Einstein first obtained such an
expression and many others such as Landau and Lifshitz,
Papapetrou, and Weinberg gave similar prescriptions (see Ref.
\cite{definitions}). The expressions they gave are called
energy-momentum complexes because they can be expressed as a
combination of $T_{i}^{k}$ and a pseudotensor, which is
interpreted to represent the energy and momentum of the
gravitational field. These complexes have been heavily criticized
because they are non-tensorial, i.e. they are coordinate
dependent.  For the Einstein, Landau-Lifshitz, Papapetrou, and
Weinberg (later ELLPW) energy-momentum complexes, one gets
physically meaningful results only in "Cartesian coordinates" (see
Ref. \cite{Cartesian,Moller}). Because of this drawback, many
others, including M\o ller \cite{Moller}, Komar\cite{Komar}, and
Penrose\cite{Penrose}, have proposed coordinate independent
definitions. Each of these, however, has its own drawbacks (for
further discussion see Ref. \cite{problems}).

    Recently, the energy momentum complexes of ELLPW have shown
promise and may prove to be more useful than the others.  Much
work has been produced in the past decade showing the these
complexes give meaningful results for many well known metrics (
Ref. \cite{results,VirKN,VirFlat}). Also, Aguirregabiria {\em et
al.}\cite{Aguirr} showed that they coincide for all Kerr-Schild
class metrics which include the Schwarzschild,
Reissner-Nordstr\"{o}m, Kerr-Newman, Vaidya, Dybney {\em et al.},
Kinnersley, Bonnor-Vaidya, and Vaidya-Patel spacetimes (for
references see in \cite{spacetimes}).

    In this Letter, we will calculate, using the energy momentum
complexes of ELLPW, the energy and momentum densities of the
metric given by Bonnor\cite{Bonnor} describing a stationary beam
of light. First we will show that the metric is of Kerr-Schild
class, and then we will use the results given by Aguirregabiria
{\em et. al.} to compute its energy and momentum densities.
Finally, we will discuss the physical implications of our
calculations.

    We use the convention that Latin indices take values from $0$
to $3$ and Greek indices take values from $1$ to $3$, and take
units where $G=1$ and $c=1$.

%========================================================================
\section{Kerr-Schild Class Metrics}

A metric is said to be of Kerr-Schild class if it can be written
in the form

\begin{equation}
g_{ab} = \eta_{ab} - 2Vl_{a} l_{b}
\end{equation}

where $\eta$ is the Minkowski metric, $V$ is a scalar field, and
$l$ is a null, geodesic, and shear free vector field in the
Minkowski spacetime. These three properties of $l$ are
respectively expressed as

\begin{eqnarray}
\eta^{ab} l_{a} l_{b}&=&0\\
\eta^{ab} l_{i,a} l_{b}&=&0\\
(l_{a,b} + l_{b,a})l^{a}_{\ ,c} \eta^{bc} - (l^{a}_{\ ,a})^{2}&=&0
\end{eqnarray}

    Aguirregabiria {\em et. al.}\cite{Aguirr} showed that for general Kerr-Schild class
metrics the ELLPW energy-momentum complexes coincide up to raising
and lowering of indices by the Minkowski metric.  Later,
Virbhadra\cite {KSV} clarified that the vector field $l$ need only
be null and geodesic for the ELLPW complexes to coincide.
    These energy momentum complexes for any Kerr-Schild class metric
are given by

\begin{eqnarray}
\Theta_{i}^{\ k}&=&\eta_{ij} L^{jk}\\
L^{ik}=\Sigma^{ik}&=&W^{ik}=\frac{1}{16\pi}\Lambda_{\ \ \ \
,lm}^{iklm}
\end{eqnarray}

where $\Theta$, $L$, $\Sigma$, and $W$ are the Einstein,
Landau-Lifshitz, Papepetrou, and Weinberg energy momentum
complexes respectively and

\begin{equation}
\Lambda^{ikpq} = 2V (\eta^{ik} l^{p} l^{q} + \eta^{pq} l^{i} l^{k}
- \eta^{ip} l^{k} l^{q} - \eta^{kq} l^{i} l^{p}).
\end{equation}

$\Theta_{0}^{0}$, $L^{00}$, $\Sigma^{00}$, and $W^{00}$ represent
total energy density and $\Theta_{\alpha}^{0}$, $L^{0\alpha}$,
$\Sigma^{0\alpha}$, and $W^{0\alpha}$ represent momentum density
in the $x^{\alpha}$ direction.

\section{The Bonnor Metric}

The Bonnor metric describing a stationary beam of light flowing in
the $Z$ direction is given by the line element\cite{Bonnor}

\begin{equation}
ds^{2} = -dx^{2} - dy^{2} - (1-m)dz^{2} - 2m dz dt + (1+m)dt^{2}
\end{equation}

where $m$ is a function of $x$ and $y$,

\begin{eqnarray}
\nabla^{2} m &=& 16 \pi \rho \label{eq:one}\\
\rho = -T_{3}^{3} &=& -T_{3}^{0} = T_{0}^{3} = T_{0}^{0}
\label{eq:two}
\end{eqnarray}

and $T_{a}^{b}$ is the energy momentum tensor.

The Bonnor metric can be rewritten as

\begin{equation}
ds^{2} = dt^{2} - dx^{2} - dy^{2} - dz^{2} - m(dt - dz)^{2}.
\end{equation}

This is in form required of a Kerr-Schild class metric with

\begin{eqnarray}
 V &=& m/2 \\
 l_{t} &=& 1\\
 l_{z} &=& -1
 \end{eqnarray}

Both components of $l$ are constant so $l$ is trivially geodesic
and shear free.  It can be easily shown to be null which proves
that the Bonnor metric is of Kerr-Schild class.

We can then easily compute the energy momentum complexes of ELLPW
for the metric.  We note that, in general $\Lambda$ exhibits the
following symmetries

\begin{equation}
\Lambda^{ikpq}=\Lambda^{kiqp}=\Lambda^{pkiq}
\end{equation}

Then, we calculate the required, non-vanishing components of
$\Lambda$ to be

\begin{eqnarray}
\Lambda^{0011} &=& \Lambda^{0022} = m \\
\Lambda^{0033} &=& 2m \\
\Lambda^{0311} &=& \Lambda^{0322} = m\\
\Lambda^{0303} &=& -2m
\end{eqnarray}

From here we calculate the energy-momentum complexes.  The
non-vanishing components are, as expected, total energy density
and momentum density in the $z$ direction. These both are given by

\begin{equation}
L^{00}=L^{03}=\frac{1}{16\pi} (m_{,xx} + m_{,yy})
\end{equation}

Then, recalling Equations (9) and (10)

\begin{equation}
L^{00}=L^{03}= \rho = T_{0}^{0} = T_{0}^{3}
\end{equation}

\section{Discussion}
    This result is what would be expected from purely physical
arguments. The energy and momentum densities calculated from the
complexes of ELLPW coincide and are equal to the energy and
momentum density components of $T_{i}^{k}$.

    This simple result supports the Cooperstock hypothesis\cite{Cooperstock}
which states that energy is localized to the region where the
energy-momentum tensor is non-vanishing.  This hypothesis would
imply that there is no energy-momentum contribution from the
"vacuum" regions of spacetime.  If true, the hypothesis would have
broad implications.  For instance, a quantum theory of gravity
requires that the gravitons which make up the field carry energy.
Also, the hypothesis suggests that gravitational waves have no
energy and that current attempts to detect these waves by
detecting their energy will be fruitless.

    In summary, we have calculated the energy and momentum for a stationary
beam of light and found that the results are physically meaningful
and support the Cooperstock hypothesis.  Certainly the questions
of energy and momentum localization in General Relativity as well
as the Cooperstock hypothesis are far from resolved, but the
Bonnor metric provides one more example where the energy momentum
complexes of ELLPW provide, for a physical metric, meaningful
results that support the hypothesis.

\begin{acknowledgments} I am grateful to K.S. Virbhadra for his
guidance, to Duke University for its support, and also to the
referee for helpful suggestions.
\end{acknowledgments}

\end{document}